\documentclass[%
 aip,
 amsmath,amssymb,
 reprint,%
]{revtex4-1}

\usepackage{graphicx}
\usepackage{dcolumn}
\usepackage{bm}
\usepackage{epstopdf}
\usepackage{algorithm}
\usepackage{algpseudocode}
\usepackage[breaklinks]{hyperref} 
\hypersetup{colorlinks=true, allcolors=blue}
\usepackage{amsmath,amsthm,amssymb}
\usepackage{float}
\usepackage{epsfig}
\usepackage{mathrsfs}
\usepackage{multirow}
\usepackage[all]{xy}
\usepackage{pbox}
\usepackage{verbatim}
\usepackage{braket}
\usepackage{mathtools}
\usepackage{tikz}
\usepackage{xcolor}
\usepackage{xfrac}
\usepackage{cleveref}

\usepackage[caption=false]{subfig}
\usepackage{float}

\usepackage{graphicx}
\usepackage{dcolumn}
\usepackage{bm}
\usepackage{amsmath}
\usepackage{mathrsfs}
\usepackage{amsfonts}
\usepackage{amssymb}
\usepackage{graphicx}
\usepackage{amsthm}

\usepackage[utf8]{inputenc}
\usepackage[T1]{fontenc}
\usepackage{mathptmx}
\usepackage{etoolbox}

\makeatletter
\def\@email#1#2{%
 \endgroup
 \patchcmd{\titleblock@produce}
  {\frontmatter@RRAPformat}
  {\frontmatter@RRAPformat{\produce@RRAP{*#1\href{mailto:#2}{#2}}}\frontmatter@RRAPformat}
  {}{}
}%
\makeatother

\begin{document}


\title[Wigner Cat Phases]{Wigner Cat Phases: A finely tunable system for exploring the 
transition to quantum chaos}
\author{M. S\"uzen}
\email{mehmet.suzen@physics.org}
\address{Assia, CY 5561, Cyprus}
\address{American Physical Society, Member, College Park, MD}

\date{\today}

\begin{abstract}
A composite quantum mechanical system consisting of a single frozen qubit 
and a fully thermalized chaotic system is investigated. Induced 
localization is tunable via the subspace dimension ratio, which serves as 
an asymmetry parameter. The system exhibits Wigner-Dyson 
level spacing statistics without tuning, indicative of quantum 
chaos and a thermal state. As the tuning parameter is reduced 
and fewer states are selected, the nearest-neighbor 
level spacing distribution develops heavier tails but does not 
fully reach Poisson statistics. In these heavy-tailed regimes, 
termed Wigner cat phases, bimodality, or 'cat-ears', 
is observed across the eigendensity and in measures of eigenstates 
via participation ratio distributions. We demonstrate that this 
subspace-induced localization mechanism can be characterized 
by a mixed Gaussian Orthogonal Ensemble (mGOE), 
which selects different-sized matrices around a given finite matrix 
size to optimize statistics and computation speed. These phases are not tied 
exclusively to the mixed ensemble, nor are they an artifact of the mGOE's numerical 
construction, as verified by randomizing state permutations on the 
composite Hamiltonian independent of this ensemble. We also note 
the implications of the mGOE for quantum graphs.
\end{abstract}

\maketitle


\section{Introduction}\label{sec:intro}

The correspondence between classical and quantum mechanical descriptions 
of systems exhibiting classical chaos has long fascinated 
physicists \cite{berry87, martinis87, ford92, gutzwiller92, heller93}. 
Defining and quantifying quantum chaos remains a central 
topic \cite{berry77, chirikov88, zurek95, nakamura94, chirikov95, 
stockmann09, santos12, gutzwiller13, wimberger14, haake18, heller18}. 
Among these, the Bohigas-Giannoni-Schmit (BGS) conjecture \cite{bohigas84} 
stands out as a foundational result: it predicts that quantum systems 
with classically chaotic dynamics exhibit level spacing statistics 
governed by Wigner-Dyson ensembles. These invariant random matrix ensembles
describe quantum systems whose classical counterparts are chaotic and 
ergodic \cite{wishart28, wigner51, wigner55, wigner57, wigner58, wigner67, 
dyson62a, dyson62b, dyson62c, dyson62d, dyson62e, mehta04, tao, potters20}. 
This connection is commonly referred to as quantum chaology (originally 
introduced by Sir Michael Berry \cite{berry89}), though we use the term 
quantum chaos for consistency with the broader literature. In prior work, 
Berry's conjecture \cite{berry77, berry20, jarzynski97} proposed that 
high-energy eigenstates are described by wavefunctions drawn from a Gaussian 
random ensemble of plane waves. Parallel to this school of thought, 
quantum graphs have been shown to provide an alternative framework to study spectral and 
nodal domain statistics when detecting quantum chaos. Using connectivity 
matrices with bond lengths \cite{kottos97, kottos99, blum02} to describe 
quantum chaos, alongside descriptions of scarring phenomena \cite{heller84, 
kaplan26}, constitutes an active area of research from a quantum chaology 
perspective.

Connections between the BGS conjecture and the emergence of quantum ergodicity are 
established through random matrix theory, particularly via the Eigenstate 
Thermalization Hypothesis (ETH) \cite{jdeutsch91, srednicki94, srednicki95, 
srednicki96, srednicki99}, as reviewed in \cite{dalessio16, deutsch18}. In this 
direction, recent studies using the Loschmidt echo and out-of-time-order correlators 
(OTOC) have advanced our understanding \cite{foini19, papalardi22, 
foini25}. Recent work also builds foundational connections between the ETH and the 
BGS conjecture \cite{weiden25}. There is a growing interest in quantum 
chaos within quantum information theory and its applications \cite{cotler17, magan18, 
ali20, altland24, borner24, anand24, googleq25}, especially regarding the advancement of quantum 
chaos detection protocols \cite{das25}. Moreover, multi-parametric ensembles have 
been employed to describe non-ergodic regimes \cite{buijsman19, rao21, rao24, 
shekhar23, shekhar25a, shekhar25b}, leveraging local spectral and eigenfunction 
statistics. 

In Section \ref{sec:comp}, we introduce a simple composite system that 
can be used as a testbed for localization studies. The system is 
experimentally motivated, and we introduce the concept of selective 
access to subspace dynamics. Similar experimental and algorithmic 
procedures have recently been studied in quantum information and 
control. Our main results identify bimodal states corresponding to the 'Wigner cat 
phases'. These bimodal states are demonstrated via exact 
diagonalization, depicting an elementary bimodality in the eigendensity 
and eigenstate statistics. In Section \ref{sec:mes}, 
the mixed ensemble is used to study the spectral statistics of 
the composite system in depth, and the variation of the asymmetry parameter 
is extensively analyzed. We present our results using canonical spectral 
analysis of localization. In Appendix \ref{sec:mixed}, we define this 
class of invariant random matrix ensembles, a mixed Wigner-Dyson 
ensemble, which remains distinct from the Brownian ensemble \cite{shekhar23}.  
We conclude the study in Section \ref{con}; our framework 
provides a continuous description of the transition to ergodicity 
within a single ensemble, characterized by 
the asymmetry parameter.

Additionally, in Appendix \ref{app:per}, we discuss how the tiling 
of spectra constructs the mGOE as an equivalence to the composite system.  
In Appendix \ref{app:bimodality}, we show why bimodality occurs in this 
composite system analytically using the ratio of the second and fourth moments.

\section{Composite System with Frozen Qubit} \label{sec:comp}

The composite system considered here consists of a frozen qubit 
and an $L$-state quantum chaotic system. We compose two finite 
Hilbert spaces \cite{nielsen10, carroll21}, where $\cong$ denotes 
isomorphic equivalence:

\begin{equation}
 \mathcal{H}_{1} \cong \mathbb{C}^{2}, \qquad \mathcal{H}_{2} \cong \mathbb{C}^{L}.
\end{equation}

The composite Hilbert space is given by \cite{carroll21}:

\begin{equation}
 \mathcal{H} = \mathcal{H}_{1} \otimes \mathcal{H}_{2}.
\end{equation}

We can express the total energy operators in matrix form,  
$H_{1} \in \mathbb{C}^{2 \times 2}$ and $H_{2} \in  \mathbb{C}^{L \times L}$, 
for the frozen qubit and the fully thermalized chaotic system, respectively. 
Given that $\mathbb{I}_{2} \in \mathbb{C}^{2 \times 2}$ 
and $\mathbb{I}_{L} \in \mathbb{C}^{L \times L}$ are identity 
matrices, and $H_{c}$ represents the coupling energy, the
total Hamiltonian of the composite system is governed by \cite{carroll21}:

\begin{equation} 
H = H_{2} \otimes \mathbb{I}_{2} + \mathbb{I}_{L} \otimes H_{1} + H_{c}. 
\end{equation}

In the case of a frozen qubit, by definition, $H_{1}=\bf{0}_{2 \times 2}$ and it lacks 
coupling to the internal degrees of freedom of the chaotic system ($H_{c}=\bf{0}_{2L \times 2L}$). 
Consequently, the composite Hamiltonian, described in the $\mathbb{C}^L \otimes \mathbb{C}^2$ basis, 
simplifies to:

\begin{equation}
H = H_{2} \otimes \mathbb{I}_{2} = \begin{pmatrix}
H_{2} & \bf{0} \\
\bf{0} & H_{2} 
\end{pmatrix}.
\end{equation}

Because $H_{2}$ represents a quantum chaotic system in a thermalized
state, the BGS conjecture \cite{bohigas84} justifies modeling 
$H_{2}$ as a random matrix from the Wigner-Dyson family, sampled from 
the Gaussian Orthogonal Ensemble (GOE).

\begin{figure}[ht!]
  \centering
  \includegraphics[width=0.65\columnwidth]{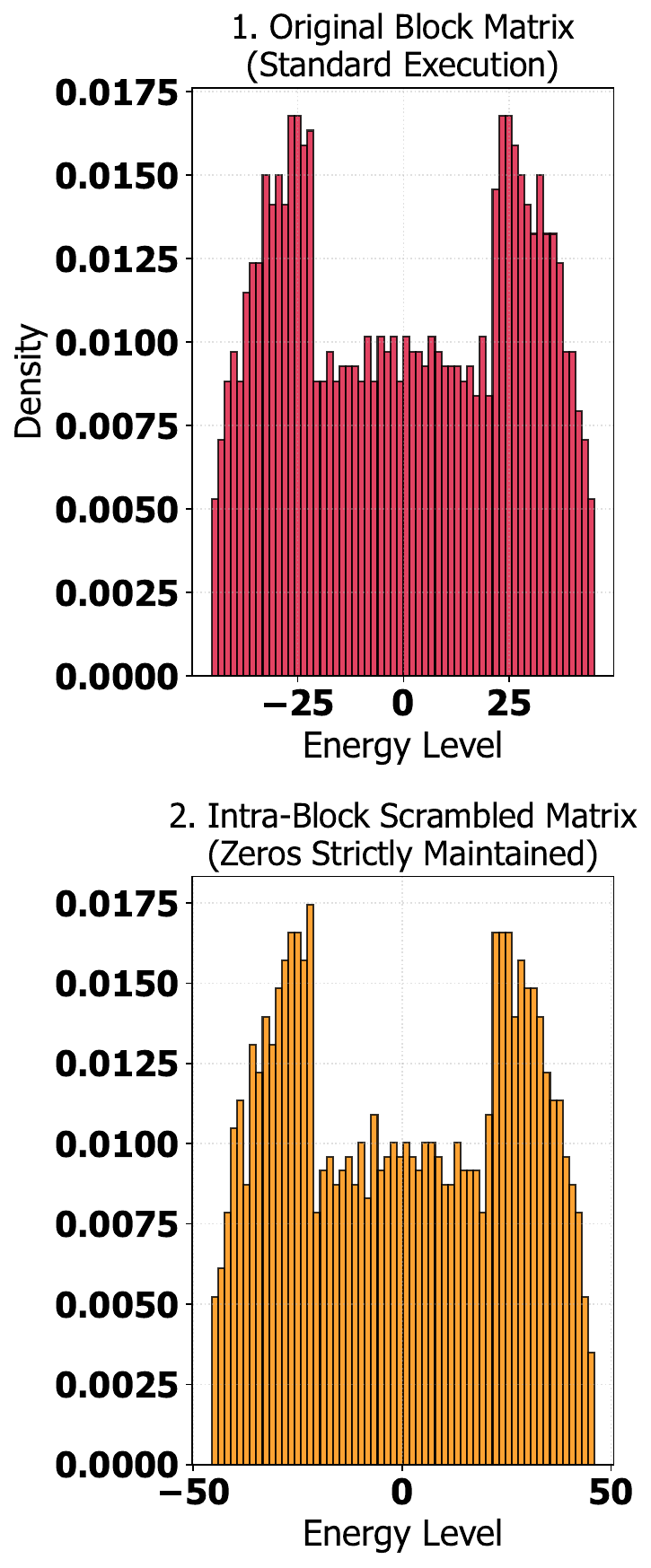}
  \caption{The eigendensity with bimodal shape direct and with 
           permutation scrambling at $\mu=0.7$ and $N=1500$. This is 
           a direct computation of the composite system's eigendensity without
           any matrix ensemble. We have shown by the permutation matrix scrambling  
           the bimodality ("cat-ears") aren't depend on the underlying eigenvalue 
           solver, rather the property of sub-selection on the composite system. 
           We should make sure that exact diagonalization is not sorting 
           the spectra automatically.}
  \label{fig:cat0}
\end{figure}

\subsection{Selective Access to Subspace Dynamics}

We restrict the spectrum of $H$ by truncating the eigenenergies at a 
reference number of states $N$, defined such that $\mu \cdot N = L$, where 
$N \in \mathbb{Z}^{+}$. This procedure corresponds to 
exploring the subspace of an $N$-state system.

The spectrum of the composite operator $H$ to be truncated at the $N$-th 
eigenenergy, alongside its individual component $H_{2}$, is expressed using an 
unsorted, random enumeration of spectral values as:

\begin{eqnarray}
\sigma(H_{2})   & = & \{ \lambda_{1}, \dots, \lambda_{L} \}, \\
\sigma(H)       & = & \{ \lambda_{1}, \dots, \lambda_{L}, \lambda_{1}, \dots, \lambda_{L}  \}, \\ 
\sigma^{\text{sub}}(H) & = & \{ \lambda_{1}, \dots, \lambda_{L}, \lambda_{1}, \dots, \lambda_{N-L}  \}. \label{comp:eigen}
\end{eqnarray}

The experimental realization of many-body localization (MBL) remains an active area of research, 
with physical platforms such as superconducting qubits \cite{xu18, guo21, liu26} representing key progress.
In this context, truncation at different values of $\mu$ in our composite system relates to 
the system's effective degree of localization. The subspaces obtained by truncation physically 
decouple chaotic degrees of freedom from ordered states due to the asymmetry imposed by 
the quantum composition rule upon selection. This behavior is reminiscent of an adiabatic or 
Born-Oppenheimer type separation \cite{born27}. Analogous physical truncations, subspaces, 
and state-merging methodologies are routinely utilized in recent studies of single magnon 
spins \cite{xu23} and frozen nuclear spin baths \cite{mateusz20}, with practical  
utility in designing fault-tolerant quantum memories \cite{brav24}. Furthermore, 
schemes utilizing state freezing and selective access to subspace dynamics have demonstrated 
utility in enhancing the scalability of quantum optimization 
algorithms \cite{ayan23} through the restriction of physical states, while related 
mechanisms of frozen coherence have been extensively explored to preserve protected 
quantum states \cite{bromley15, fu24}. The tuning of $\mu$ occurs naturally
in the physical system due to topological or physical constraints, such as
hotspot skipping or related filtering mechanisms \cite{ayan23}.

To systematically verify that the emerging localization behavior is not an artifact of 
state enumeration or a consequence of the internal sorting routines in exact diagonalization, 
we introduce an independent basis-scrambling transformation. 
Let $\sigma_1, \sigma_2 \in S_{L}$ be two distinct, randomly chosen 
elements of the symmetric permutation group of degree $L$. We 
construct their corresponding orthogonal permutation 
matrices, $P(\sigma_1)$ and $P(\sigma_2)$, defined by 
the matrix elements:
\begin{equation}
P_{jk}(\sigma_m) = \delta_{j, \sigma_m(k)} \quad \text{for } m \in \{1,2\}.
\label{eq:permutation_elements}
\end{equation}

We define the transformed, intra-scrambled composite 
Hamiltonian $H_{\text{scramble}}$ via the action 
of the decoupled direct-sum permutation operator 
$\mathcal{P} = P(\sigma_1) \oplus P(\sigma_2)$:

\begin{align}
H_{\text{scramble}} &= \mathcal{P} H \mathcal{P}^T \nonumber \\
&= \begin{pmatrix} 
P(\sigma_1) & \mathbf{0}_{L} \\ 
\mathbf{0}_{L} & P(\sigma_2) 
\end{pmatrix} 
\begin{pmatrix} 
H_2 & \mathbf{0}_{L} \\ 
\mathbf{0}_{L} & H_2 
\end{pmatrix} 
\begin{pmatrix} 
P(\sigma_1)^T & \mathbf{0}_{L} \\ 
\mathbf{0}_{L} & P(\sigma_2)^T 
\end{pmatrix} \nonumber \\
&= \begin{pmatrix} 
P(\sigma_1) H_2 P(\sigma_1)^T & \mathbf{0}_{L} \\ 
\mathbf{0}_{L} & P(\sigma_2) H_2 P(\sigma_2)^T 
\end{pmatrix}.
\label{eq:scramble_algebra}
\end{align}

Because each individual permutation matrix is strictly 
orthogonal ($P(\sigma_m) P(\sigma_m)^T = \mathbb{I}_{L}$), the global block 
operator $\mathcal{P}$ is unitary over the composite space. Consequently, 
the global spectrum is invariant under this transformation:
\begin{equation}
\text{Spec}(H_{\text{scramble}}) \equiv \text{Spec}(H).
\label{eq:spectral_identity}
\end{equation}

This permutation approach confirms that the truncation isolates 
the localization of states, driven by the physics of the composition rather than the arbitrary 
enumeration of spectra. This characteristic behavior remains invariant even when $P(\sigma_1)=P(\sigma_2)$.

Utilizing a direct solver and a permutation matrix to scramble the composite system before truncation, 
Figure \ref{fig:cat0} demonstrates the bimodal clustering of localized states deviating 
from the baseline chaotic signature. This clear deviation is illustrated using a realization at $\mu=0.7$ 
with a truncation of $N=1500$ states for the chaotic sub-component directly, without relying on 
any specialized matrix ensemble to represent the composite system.
\subsection{Quantifying Localization via Eigenstates}

If a quantum system fails to achieve full thermalization, its eigenstates exhibit 
a degree of spatial or state-space localization. Within the extensive framework of 
many-body localization (MBL) \cite{kravtsov15, alet18, abanin19, sierant25}, this study focuses 
on the Inverse Participation Ratio (IPR) \cite{fyodorov}. The IPR has been studied 
extensively in the contexts of random banded matrix models \cite{evers} 
and quantum spin chains \cite{misguich}. 

We consider a normalized real-symmetric $\alpha$-th eigenstate, 
$\mathbf{\psi}_\alpha = (\psi_{\alpha, 1}, \psi_{\alpha, 2}, \dots, \psi_{\alpha, D})^T$.
The $q^{\text{th}}$-order participation ratio is defined by the intensity moments as follows:
\begin{equation}
I_{2q}(\alpha) = \sum_{k=1}^{D} |\psi_{\alpha, k}|^{2q},
\label{eq:generalized_ipr}
\end{equation}
where $D$ represents the Hilbert space dimension. 

Setting $q=2$ yields the canonical fourth-moment Inverse Participation 
Ratio, $I_4(\alpha) = \sum_{k=1}^{D} \psi_{\alpha, k}^4$. To evaluate the 
localization properties of a given set of quantum states, $I_{4}$ provides a rigorous 
quantitative measure. Ergodic (chaotic) regimes are characterized by 
complete delocalization, where $I_{4}$ displays characteristically low values governed by 
standard Gaussian Orthogonal Ensemble (GOE) statistics. 

\begin{figure}[ht!]
  \centering
  \includegraphics[width=0.95\columnwidth]{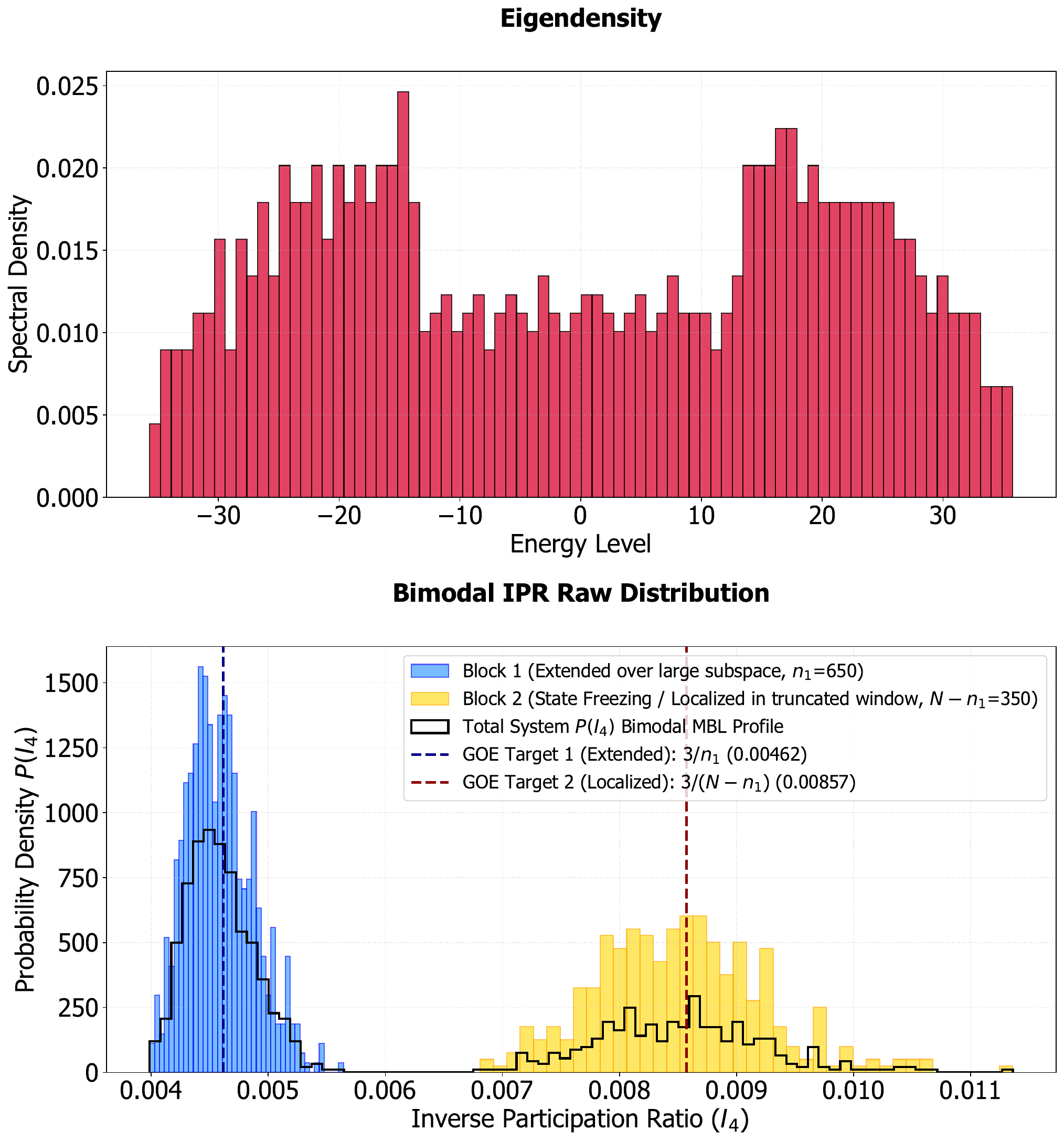}
  \caption{Wigner Phases, so called "cat-ears" explains different localizations 
   at $\mu=0.65$ and $N=1000$. (a) Shown the corresponding eigendensity with bimodal
   distribution. (b) We split the blocks at the selection, 
   and compute the distribution Inverse Participation Ratio (IPR) for selected states. 
   Resulting $P(I_4)$ distributions: We detected two distinct localization 
   showing "cat-ears".}
   \label{fig:cat1}
\end{figure}

\begin{figure}[ht!]
  \centering
  \includegraphics[width=1.0\columnwidth]{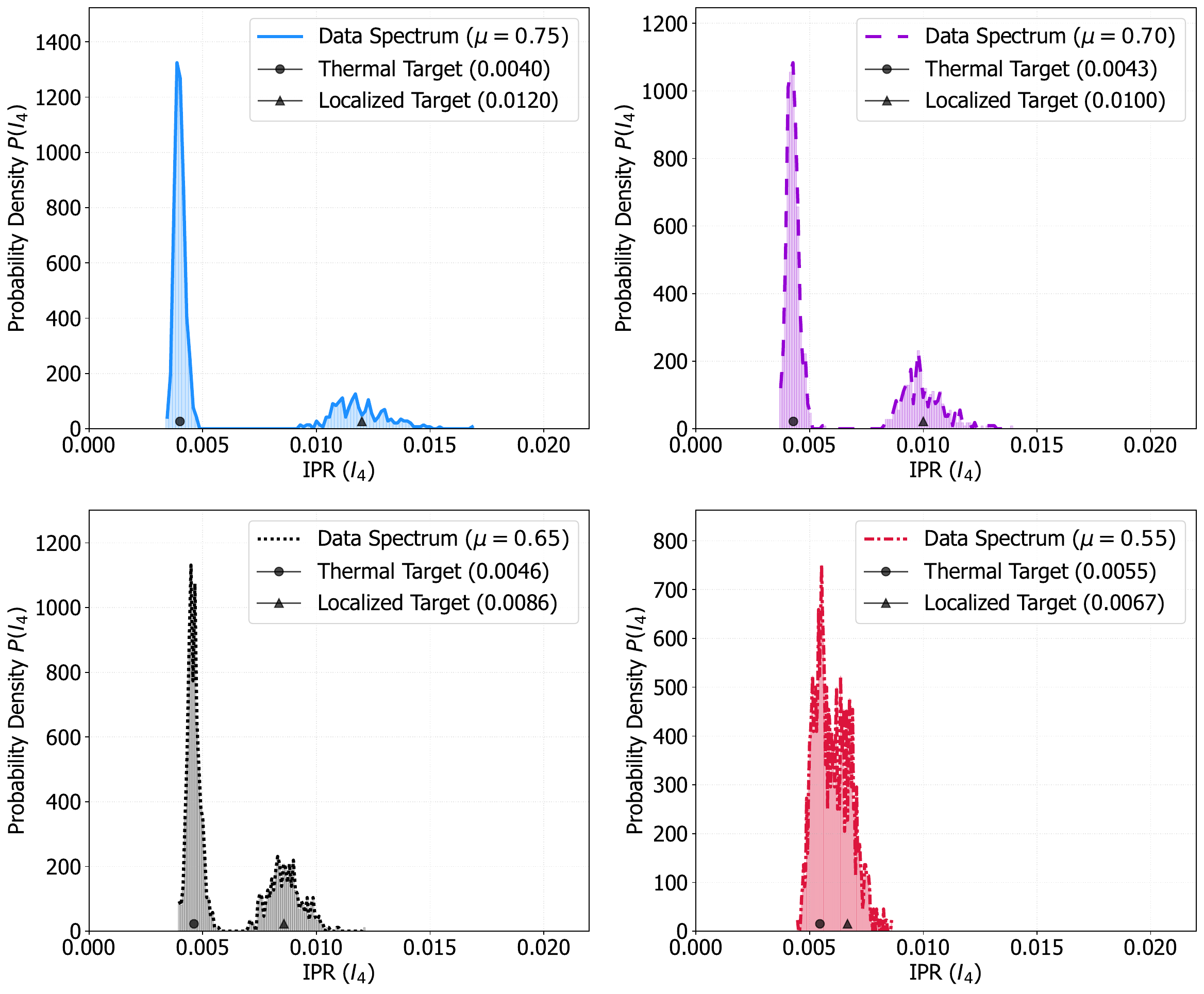}
  \caption{The changing of the distribution of Inverse Participation Ratio (IPR)
   $P(I_{4})$ for selected states for different $\mu$ values, 0.75, 0.70, 0.65 and 
   0.55. We see that "cat-ears" start to merge for lower $\mu$ values.}
   \label{fig:cat2}
\end{figure}

By computing the probability density distribution $P(I_4)$ at an asymmetry parameter 
$\mu=0.65$ with $L=1000$, we observe two distinct localization behaviors, resulting in a bimodal 
distribution. Figure \ref{fig:cat1} illustrates this bimodality for the individual diagonal matrix blocks 
post-truncation and for the total composite system post-truncation. In Figure \ref{fig:cat2}, 
this distribution profile is evaluated across varying $\mu$ values. Numerically, the state enumeration 
at the target truncation point $N$ is collected column-wise, representing a projection or subspace 
selection at the eigenstate level. For a fully delocalized state, the expected value scales inversely with 
the dimension of the active subspace:
\begin{equation}
\langle I_4 \rangle_{\text{ergodic}} \approx \frac{3}{D}.
\label{eq:ergodic_ipr_scaling}
\end{equation}

Our numerical results show close agreement with this analytical expectation value in Figure \ref{fig:cat1}. 
The separation of these subspaces under the truncation mechanism within the composite system 
supports the macroscopic "cat-ears" behavior observed in the corresponding eigendensity profiles, 
reflecting an asymmetric structural partitioning between the localized subspaces.

\section{Mixed Ensemble Spectral Statistics} \label{sec:mes}

A mixed random matrix ensemble is formulated to simulate the subspace sub-selection 
process described in the preceding section. Under this framework, sub-blocks 
are generated with varying dimensions near a nominal size $L$, and their resulting spectra 
are extended to the target selection cutoff $N$ via periodic boundary conditions using a 
cyclic list. The explicit numerical construction of this mixed-sized Gaussian Orthogonal 
Ensemble (mGOE) is detailed in Appendix \ref{sec:mixed}.

The primary task concerns the statistical identification of structural deviations 
from quantum chaos, characterizing the precise conditions and conditions under which non-ergodic profiles 
emerge. Standard canonical approaches for analyzing spectral fluctuations are applied here. 
Following a self-consistent spectral unfolding procedure, the spectral density, 
nearest-neighbor spacing distributions, and adjacent gap ratio statistics are evaluated 
across varying degrees of asymmetry.

A systematic numerical investigation is conducted to examine how tuning the mGOE parameter 
$\mu$ drives deviations from the standard limiting value $\mu = 1.0$. At this upper limit, 
the eigenvalue statistics are consistent with the predictions of the Eigenstate Thermalization 
Hypothesis (ETH), corresponding to the dynamics of a representative quantum system in the fully 
ergodic regime. A decrease in $\mu$ signifies a continuous transition toward localized states, 
corroborated by the behavior of the Inverse Participation Ratios (IPRs) analyzed in Section \ref{sec:comp}. 
Nominal block dimensions of $L = 1000$ and $L = 500$ are employed alongside an ensemble sample 
size of $M = 100$ realizations to quantify statistical uncertainties and evaluate the structural 
stability of the observed transitions.

\begin{figure}[htbp!]
  \centering
  \subfloat[]{\label{fig:s1}\includegraphics[width=0.45\columnwidth]{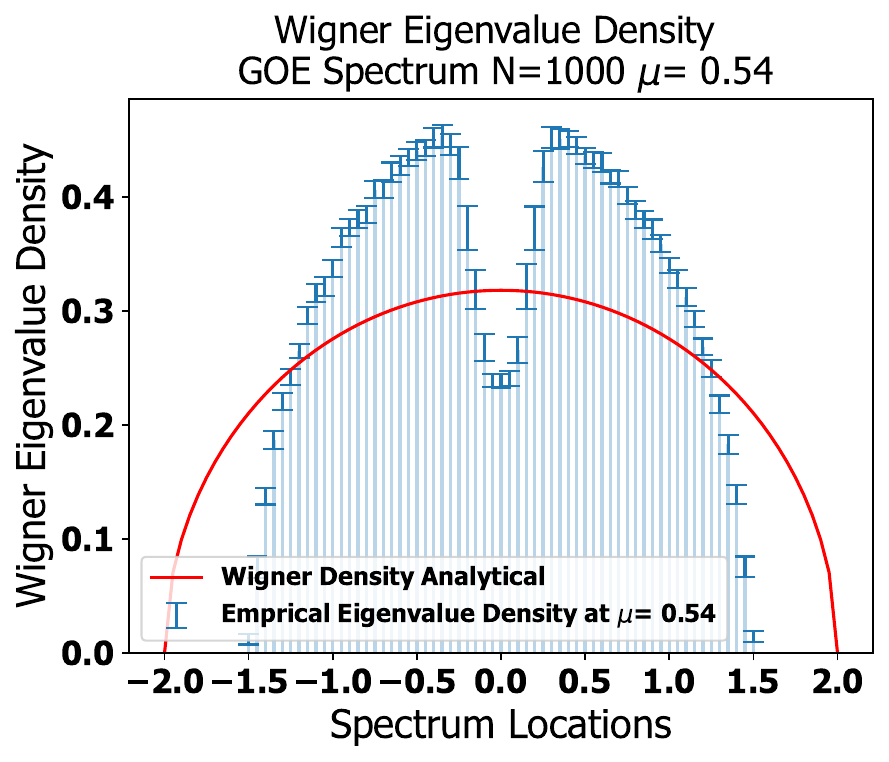}}
  \hfill    
  \subfloat[]{\label{fig:s2}\includegraphics[width=0.45\columnwidth]{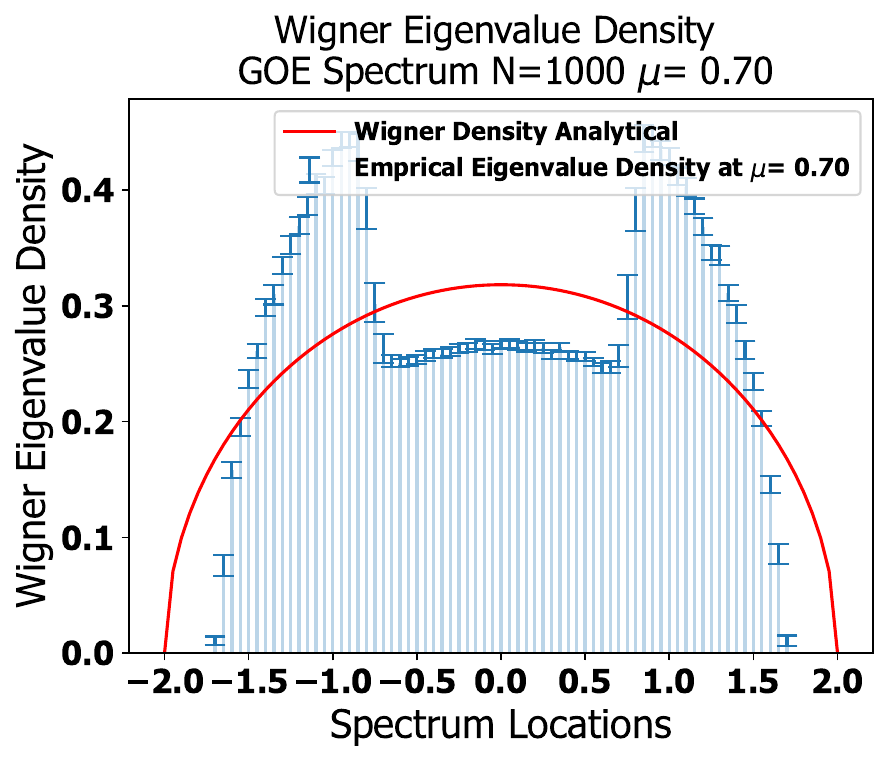}}
  \hfill    
  \subfloat[]{\label{fig:s3}\includegraphics[width=0.45\columnwidth]{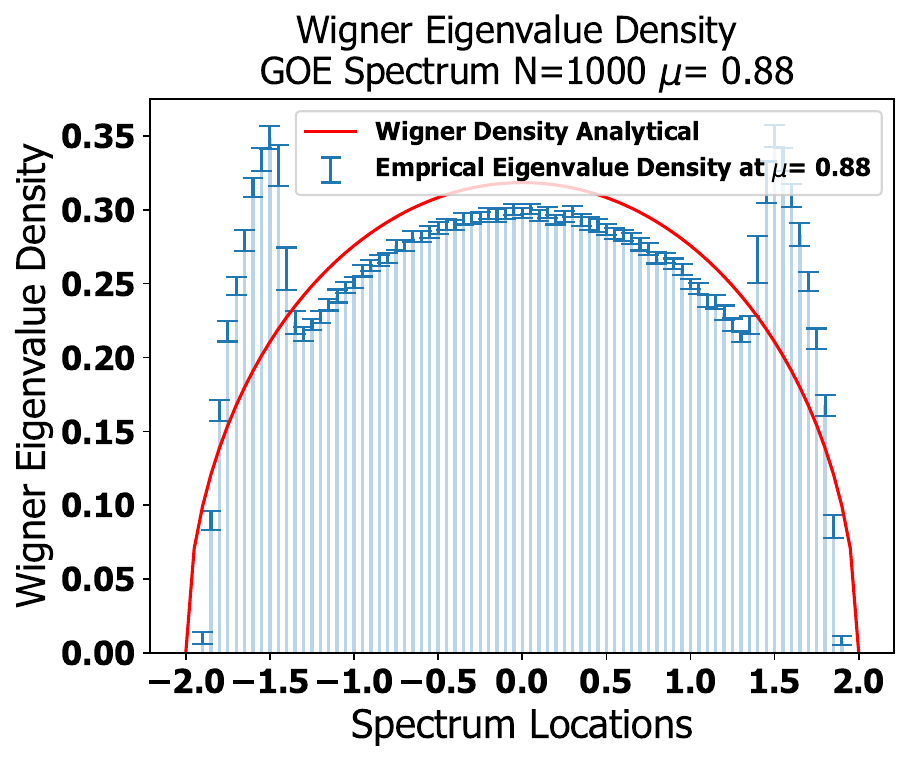}}
  \hfill    
  \subfloat[]{\label{fig:s4}\includegraphics[width=0.45\columnwidth]{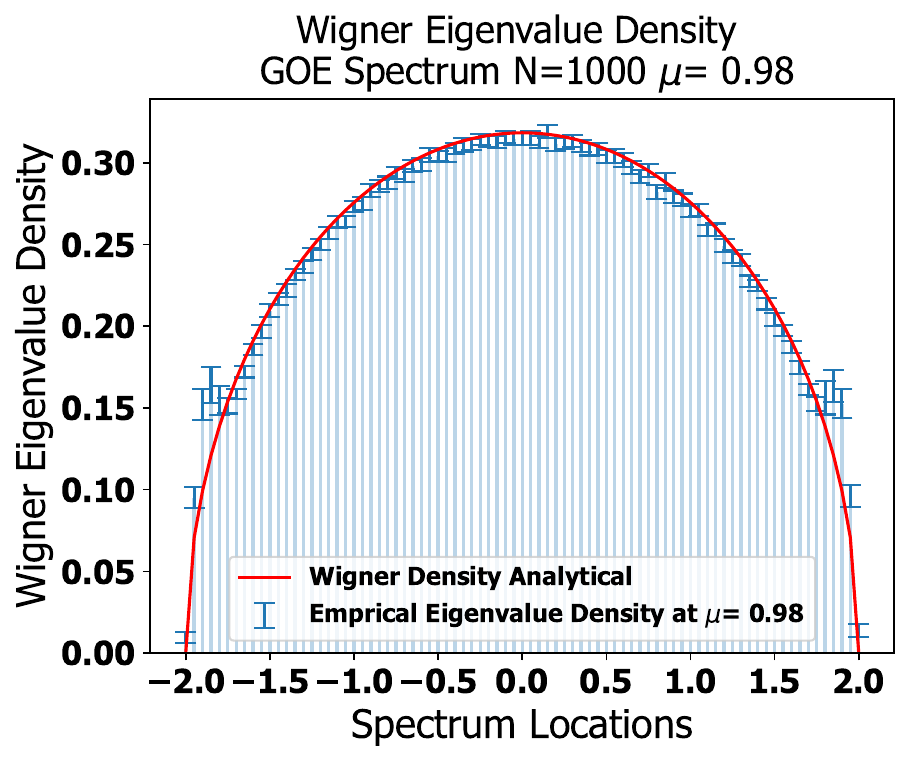}}
  \hfill    
  \caption{
  Spectral densities numerically computed for different asymmetry parameters: 
  \protect\subref{fig:s1} $\mu=0.54$, \protect\subref{fig:s2} $\mu=0.70$, \protect\subref{fig:s3} $\mu=0.88$, and 
  \protect\subref{fig:s4} $\mu=0.98$. These profiles represent the Wigner Cat Phases, identified by 
  distinct M-shaped spectral densities that show clear deviations from Wigner's semicircle law. 
  Statistical uncertainties are estimated using bootstrapped 95\% confidence intervals computed 
  over the mGOE ensemble and are displayed as error bars. The standard semicircle distribution is 
  recovered asymptotically as $\mu \rightarrow 1.0$.}
\end{figure}

To characterize the non-ergodic transition toward localization, a comprehensive dataset is 
generated using the mGOE framework across a wide range of $\mu$ values while maintaining a 
fixed ensemble size. The evolution of the system is monitored by analyzing the morphological 
variations in the spectral density, the nearest-neighbor spacing distribution $P(s)$, and the 
probability density of the adjacent gap ratios $P(r)$ as functions of the asymmetry parameter $\mu$. 
Additionally, the mean adjacent gap ratio $\langle r \rangle$ is tracked continuously across the 
entire tuning range.

\begin{figure}[htbp!]
  \centering
  \subfloat[]{\label{fig:nn1}\includegraphics[width=0.45\columnwidth]{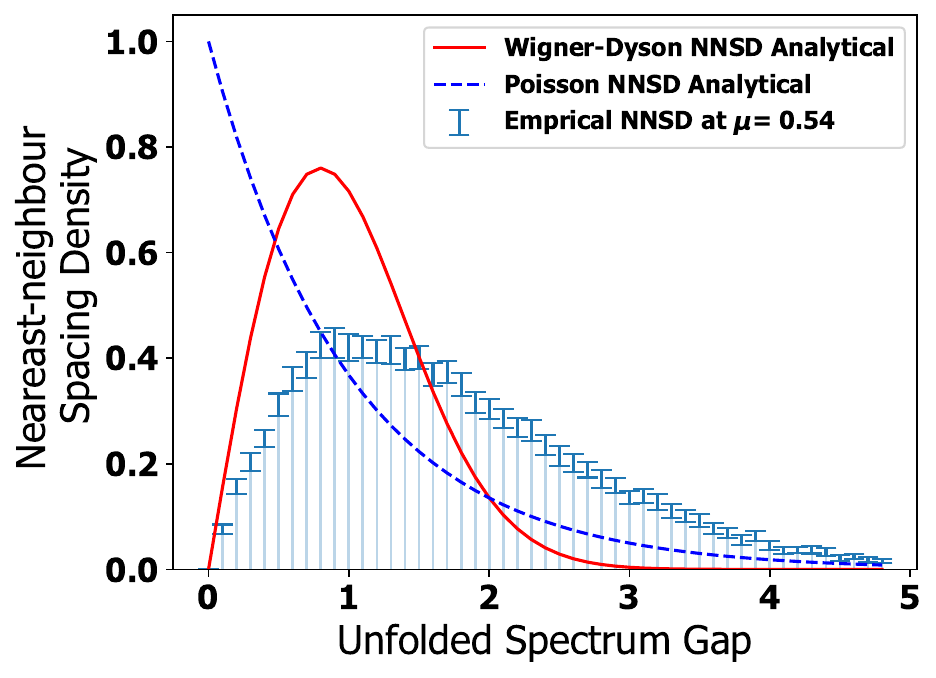}}
  \hfill    
  \subfloat[]{\label{fig:nn2}\includegraphics[width=0.45\columnwidth]{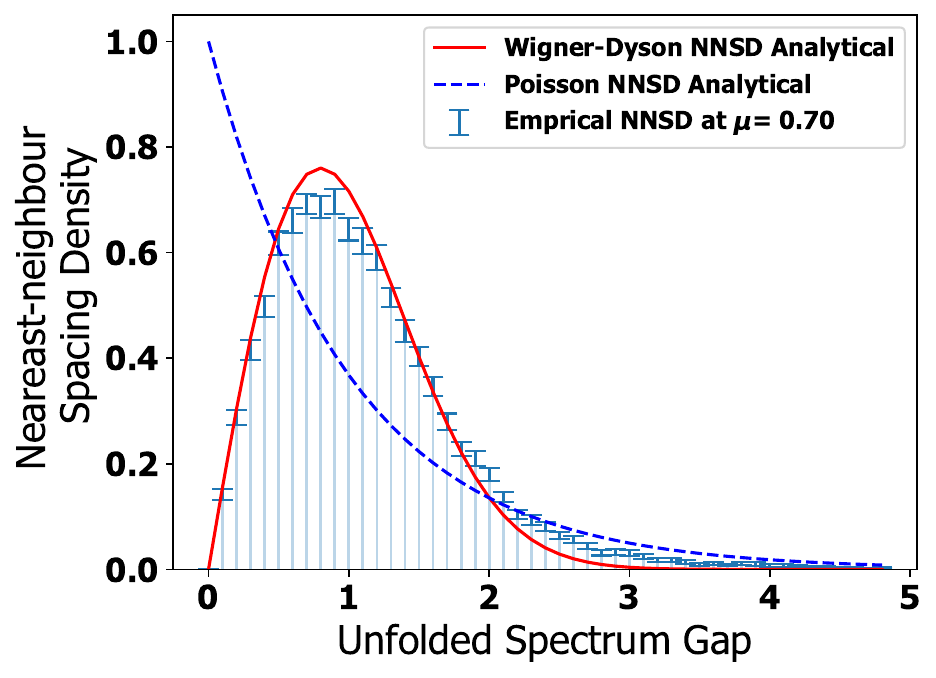}}
  \hfill    
  \subfloat[]{\label{fig:nn3}\includegraphics[width=0.45\columnwidth]{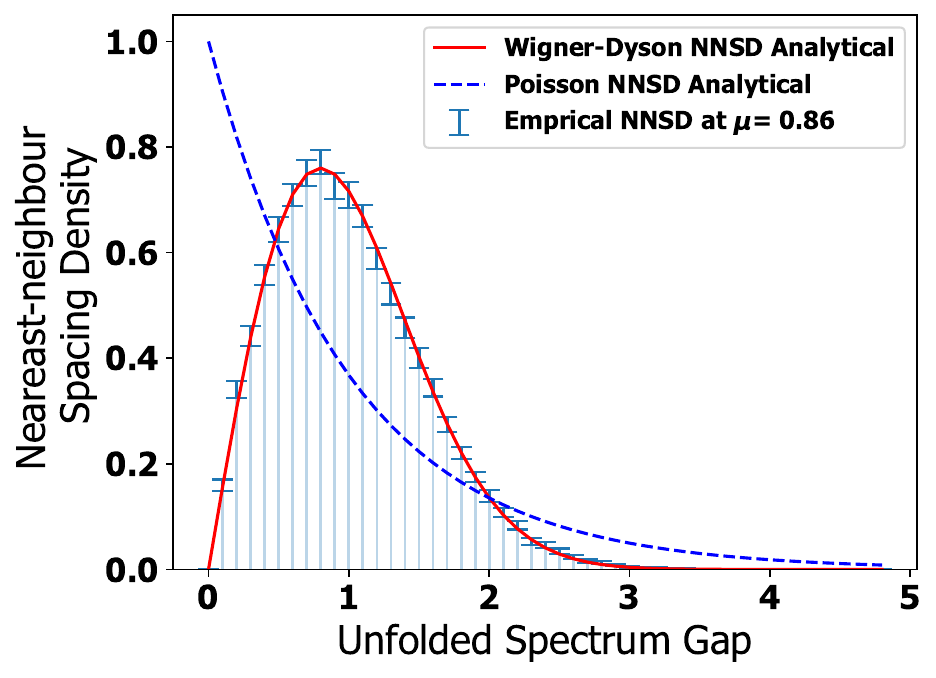}}
  \hfill    
  \subfloat[]{\label{fig:nn4}\includegraphics[width=0.45\columnwidth]{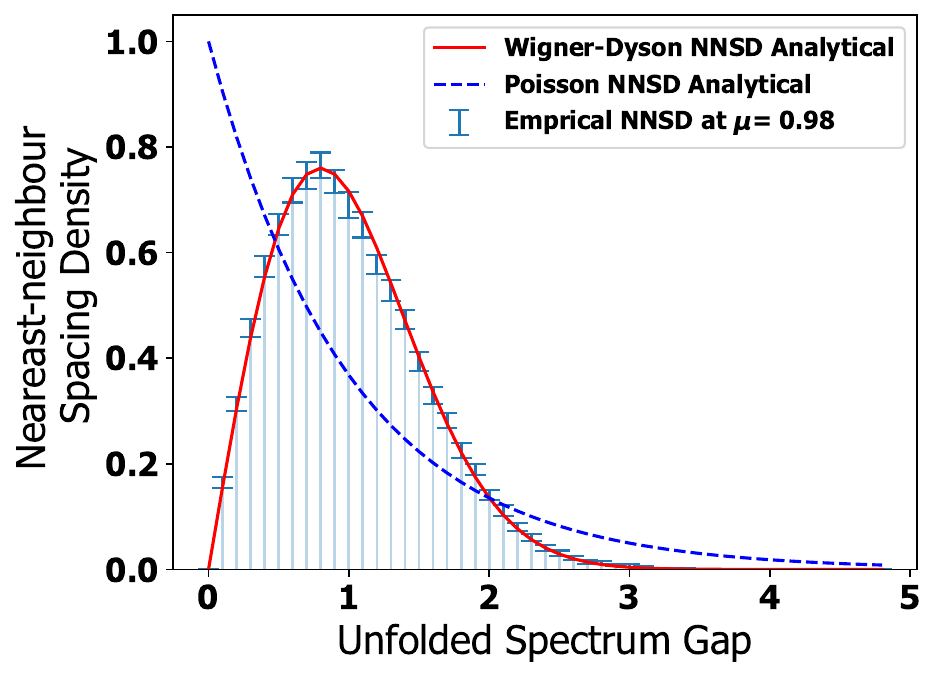}}
  \hfill    
  \caption{Nearest-neighbor spacing density $P(s)$ evaluated for various values of $\mu$, where 
  $s$ denotes the normalized spacing obtained from the unfolded spectrum: 
  \protect\subref{fig:nn1} $\mu = 0.54$, \protect\subref{fig:nn2} $\mu = 0.70$, \protect\subref{fig:nn3} $\mu = 0.86$, and 
  \protect\subref{fig:nn4} $\mu = 0.98$. For smaller values of $\mu$, the distribution shifts away from 
  the standard Wigner-Dyson form, exhibiting an increased density at small spacings alongside heavy tails. 
  Notably, a conventional Poisson distribution is not established at any value of $\mu$, indicating 
  that full integrable-like decoupling is absent. Statistical deviations from the GOE baseline are 
  quantified using bootstrapped 95\% confidence intervals (shown as error bars). As $\mu$ increases, 
  the spacing profile continuously approaches the Wigner-Dyson distribution, confirming the onset of 
  level statistics characteristic of quantum chaos.}
\end{figure}

\begin{figure}[htbp!]
  \centering
  \subfloat[]{\label{fig:ag1}\includegraphics[width=0.45\columnwidth]{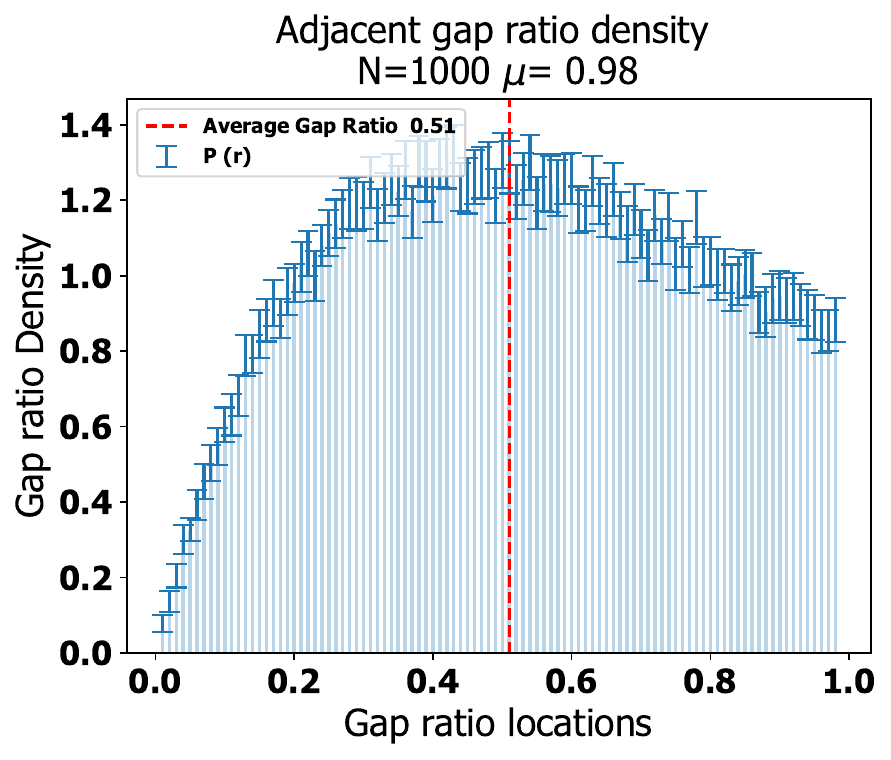}}
  \hfill    
  \subfloat[]{\label{fig:ag2}\includegraphics[width=0.45\columnwidth]{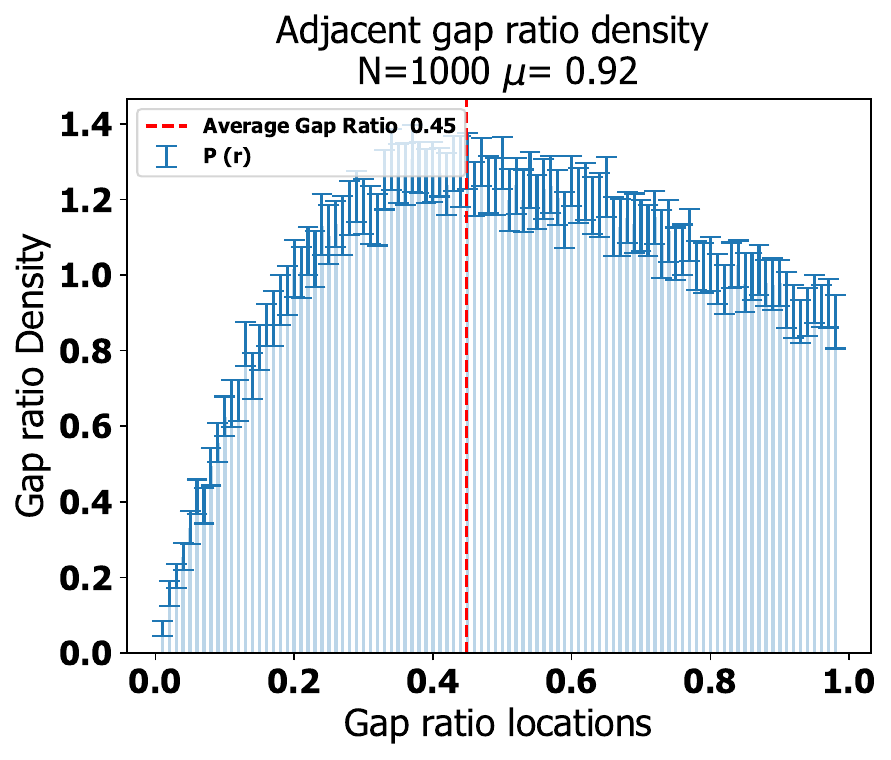}}
  \hfill    
  \subfloat[]{\label{fig:ag3}\includegraphics[width=0.45\columnwidth]{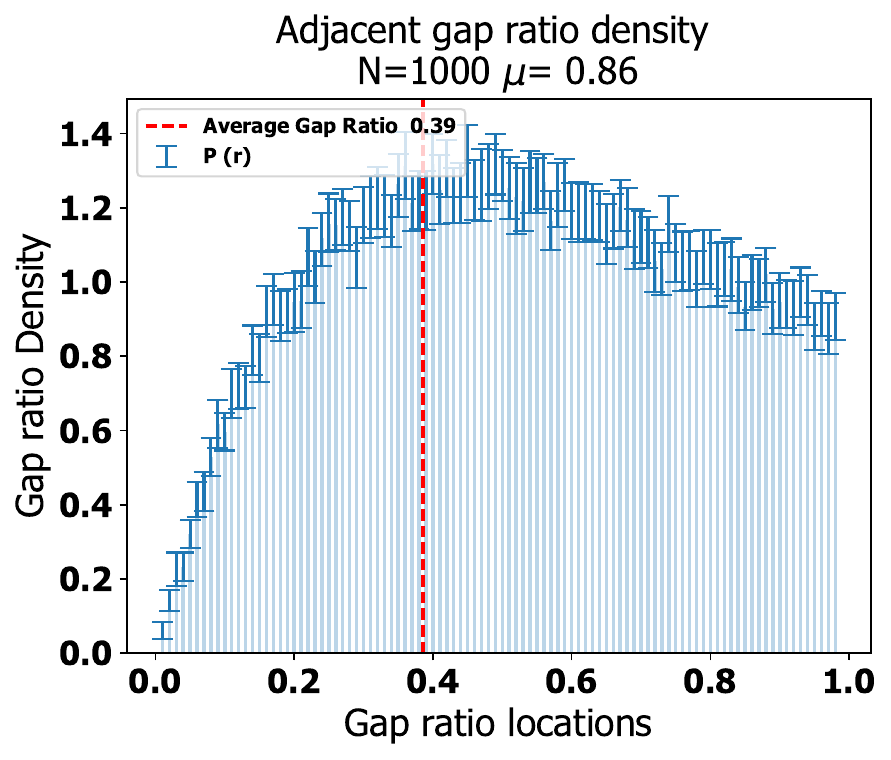}}
  \hfill    
  \subfloat[]{\label{fig:ag4}\includegraphics[width=0.45\columnwidth]{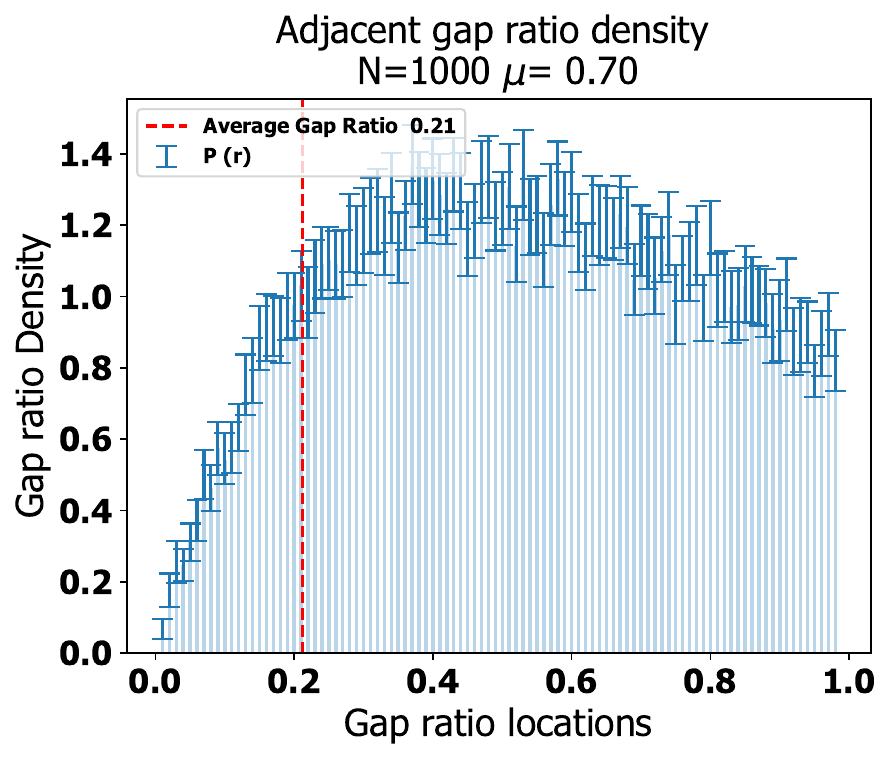}}
  \hfill    
  \caption{
  Probability density of the adjacent gap ratios $P(r)$ as a function of the gap ratio coordinate $r$, 
  with the ensemble mean values marked across different mixing parameters: \protect\subref{fig:ag1} $\mu=0.98$, 
  \protect\subref{fig:ag2} $\mu=0.92$, \protect\subref{fig:ag3} $\mu=0.86$, and \protect\subref{fig:ag4} $\mu=0.70$. 
  Uncertainties are determined across the mGOE realizations via bootstrapped 95\% confidence intervals 
  and are presented as error bars. A clear structural modification of the distribution tail is observable 
  at higher levels of asymmetry (lower values of $\mu$).
  }
\end{figure}

\subsection{Self-Consistent Spectral Unfolding}\label{sec:unf}

Spectral unfolding is applied to eliminate macroscopically varying local fluctuations and achieve 
a locally uniform spectral density on average, ensuring a reliable analysis of the fluctuation statistics. 
As established in random matrix theory literature, the calculation of nearest-neighbor spacing distributions 
can exhibit a high sensitivity to outliers or spectral edge effects in individual eigenvalues \cite{abu18}. 
To mitigate these compounding factors, the inter-quartile range (IQR) of the eigenvalue distribution is utilized 
prior to unfolding, ensuring that the resulting spectral fluctuations preserve the statistical fidelity of 
the underlying ensemble. In the fully chaotic GOE limit ($\mu = 1.0$), the unfolded spectrum shows close 
agreement with the theoretical prediction of a flat, locally uniform spacing distribution, validating the 
capacity of this procedure to isolate the intended spectral statistics.

A self-consistent optimization procedure is implemented by evaluating multiple polynomial fitting structures of varying 
degrees and selecting the specific degree that minimizes the mean fluctuation of the unfolded level spacings. 
The target objective function minimizes the mean deviation from unity. While the adjacent gap ratio distribution 
remains inherently invariant under choices of unfolding protocols, the unfolded spectrum is strictly maintained across 
all subsequent fluctuation analyses to preserve systematic consistency.

\begin{figure}[h!] 
  \centering
  \includegraphics[width=0.95\columnwidth]{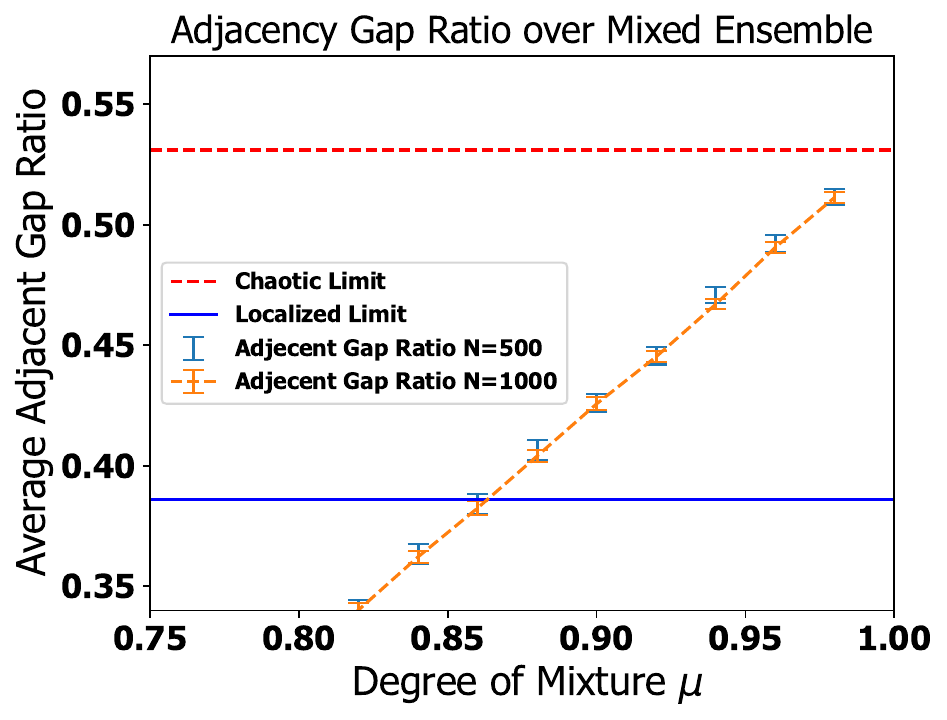}
  \caption{
    Average adjacent gap ratio $\langle r \rangle$ plotted as a function of the asymmetry parameter $\mu$, 
    illustrating the modification of localization properties. The horizontal reference lines denote the standard 
    limiting regimes: the upper bound matches full eigenstate thermalization (ETH) characterized by the 
    Wigner-Dyson (GOE) prediction ($\langle r \rangle \approx 0.530$), while the lower bound indicates conventional 
    integrability governed by Poisson statistics ($\langle r \rangle \approx 0.386$). In this system, despite the 
    numerical value of the mean ratio shifting downwards, a true Poisson distribution is not realized due to 
    the persistence of heavy-tailed statistics.
  }
  \label{fig:gsm}
\end{figure}

\subsection{Wigner Cat Phases: Spectral Densities} \label{sec:cat}

Spectral densities are numerically evaluated across a range of asymmetry parameters, 
as shown in Figures \ref{fig:s1}–\ref{fig:s4} for $\mu = 0.54, 0.70, 0.88$, and $0.98$, 
respectively. These profiles reveal the emergence of the Wigner cat phases, characterized by 
distinct M-shaped density profiles ("cat ears") that deviate significantly from Wigner's semicircle law. 
Statistical uncertainties are determined over the mGOE ensemble using bootstrapped $95\%$ 
confidence intervals and are presented as error bars. Certain error bars display asymmetry, 
reflecting the non-parametric nature of the underlying error distribution. The standard 
semicircle law is systematically recovered in the limit of weak mixing, corresponding to 
higher values of $\mu$.  
 
As the asymmetry parameter $\mu$ decreases—corresponding to an increasing degree of localization—the 
Wigner cat phases become more prominent, causing the characteristic spectral peaks to shift closer 
together. This structural reorganization arises from the interplay between eigenstate degeneracy 
and statistical randomness, generating an effective defect structure within the matrix configuration 
as described in the mGOE formulation.

\subsection{Spectral Nearest-Neighbor Spacings}\label{sec:nn}

The evolution of the nearest-neighbor spacing distribution (NNSD) is monitored continuously 
as a function of the asymmetry parameter. Spectral fluctuations are isolated using 
a self-consistent unfolding protocol combined with the truncation of spectral edge outliers \cite{ash14, abu18}. 
The ratio of consecutive level spacings (adjacent gap ratio) is analyzed across the 
transition, which is accessible within the mixed Gaussian Orthogonal Ensemble (mGOE) 
framework \cite{oganesyan07, chavda14, jisha24}. 

Nearest-neighbor spacing statistics for various values of $\mu$ are illustrated 
in Figures \ref{fig:nn1}–\ref{fig:nn4} for $\mu = 0.54, 0.70, 0.86$, and $0.98$, 
respectively. Clear deviations from the standard Wigner-Dyson distribution are observable at 
smaller values of $\mu$, where the level spacing displays a heavy-tailed profile. 
Uncertainties are computed over the mGOE ensemble using bootstrapped $95\%$ 
confidence intervals and are displayed as error bars. The standard Wigner-Dyson distribution 
is recovered at larger mixtures ($\mu \rightarrow 1.0$). Notably, a fully Poissonian distribution 
is not established even at small values of $\mu$. This indicates that the mGOE framework characterizes 
non-ergodic phases that can be continuously tuned between heavy-tailed localization and quantum chaos.

\subsection{Transitions to Quantum Chaos: Adjacent Gap Ratios}\label{sec:gap}

A diagnostics framework for this transition was introduced by Oganesyan and Huse \cite{oganesyan07}. 
Given $N$ spectral spacings $\delta_i$ obtained from sorted eigenenergies, the mean 
adjacent gap ratio $\langle r \rangle$ is defined as follows:

\begin{equation} \label{eq:agap}
\langle r \rangle = \frac{1}{N-1} \sum_{i=2}^{N} \frac{\min(\delta_i, \delta_{i-1})}{\max(\delta_i, \delta_{i-1})}.
\end{equation}

For fully localized and fully thermalized regimes, the expected value of $\langle r \rangle$ is 
given by $0.3860$ and $0.5295$, respectively \cite{oganesyan07}. It is critical to note 
that a conventional, fully localized phase is not established within our composite testbed system. 
These limiting reference values correspond to Poisson and Wigner–Dyson distributed 
energy spectra, which characterize standard integrable and chaotic quantum systems, 
respectively. The tunable ensemble presented here simulates a continuous crossover from 
eigenstate thermalization (ETH) behavior to a localized phase, corresponding to a transition from 
quantum integrability to non-integrability, and from quantum non-ergodicity to ergodicity.

Probability density distributions of the adjacent gap ratios $P(r)$, with ensemble mean values 
indicated across different mixing parameters, are shown in Figures \ref{fig:ag1}–\ref{fig:ag4} 
for $\mu = 0.98, 0.92, 0.86$, and $0.70$, respectively. Uncertainties are computed using 
bootstrapped $95\%$ confidence intervals.  

The mean values of these gap ratios provide a quantitative metric for the degree of 
localization in the quantum dynamics. The mGOE ensemble captures this crossover in a 
continuous manner, as tracked in Figure \ref{fig:gsm}. In the highly localized limit—where 
the classical analog typically exhibits integrable behavior—the mGOE retains a near-integrable signature 
characterized by persistent heavy tails rather than settling into a conventional, fully integrable structure, 
as illustrated in Figure \ref{fig:ag1}.  

\section{Conclusion} \label{con}

This study evaluates a composite quantum system featuring a frozen qubit that 
exhibits a distinct form of localization when a subset of the spectrum is 
selectively observed. This localization behavior is tunable via the asymmetry parameter 
$\mu$, which dictates the truncation threshold during the partial observation of the 
global spectrum. The bimodal nature of both the eigendensity and the eigenstate localization 
profiles is characterized by an emergent "cat-ear" structure. These numerical investigations 
are enabled by a mixed Gaussian Orthogonal Ensemble (mGOE) that models the 
composite system while offering enhanced statistical averaging capabilities.
 
Although the total composite system remains fundamentally thermalized, localization 
emerges within the observed subsystem. Such selective spectral probing, where a defined portion 
of the spectrum is isolated, can serve as a flexible testbed for non-ergodic phenomena. 
Further mapping of these transitions will benefit from detailed quantitative analyses 
extending beyond the baseline demonstration of eigenstate bimodality.

{\bf Author Contributions (CRediT):} M. S\"uzen: Conceptualization, Data curation, 
Formal analysis, Software, Validation, Visualization, Writing -- original draft, 
Writing -- review \& editing. \\ 

{\bf Data and Code Availability:} The software toolkit Leymosun \cite{suzen25ley} 
is publicly available and enables full reproducibility of the results. The 
associated dataset is also openly accessible \cite{suzen26wcp}. \\ 

{\bf Acknowledgments:} We are grateful to Y. S\"uzen for her generous support. 
The author thanks the quantum chaos community for their insightful reviews and 
constructive recommendations, which greatly helped in contextualizing the literature 
and refining the physics and presentation of this work. We also thank Devanshu 
Shekhar for kind correspondence. \\

\bibliographystyle{apsrev4-2} 
\bibliography{suzen}


\appendix

\section{Numerical Construction of the Mixed Ensemble} \label{sec:mixed}

The primary invariant ensemble representing a quantum chaotic system, as justified 
by the BGS conjecture, is the Gaussian Orthogonal Ensemble (GOE). Its numerical 
construction was detailed by Edelman and Rao \cite{edelman05}, alongside other 
generalized matrix ensembles. Here, we extend the GOE construction to a distribution 
spanning multiple matrix dimensions around a fixed reference size. 

We consider an $L \times L$ GOE matrix, which structurally corresponds to the individual 
diagonal blocks defining the composite system. A real random matrix $G_{1}$ with independent, 
identically distributed entries drawn from a normal distribution with mean zero and standard 
deviation $\sigma$ is utilized to sample a member of the GOE, denoted $A^{\text{GOE}}(L)$, 
via the symmetrization relation \cite{edelman05}:

\begin{equation} \label{eq:goe:matrix}
A^{\text{GOE}}(L) = \frac{1}{2}\left(G_1(L) + G_1^T(L)\right),
\end{equation}

where $G_{1}^{T}(L)$ denotes the transpose of $G_{1}(L)$. The sampled matrix 
$A^{\text{GOE}}(L)$ possesses the following statistical properties:
\begin{enumerate}
 \item The diagonal elements follow a normal distribution with 
       variance $\sigma^2$: $\text{diag}(A^{\text{GOE}}) \sim \mathcal{N}(0, \sigma^2)$,
 \item The off-diagonal elements follow a normal distribution with 
       variance $\sigma^2/2$: $\text{offdiag}(A^{\text{GOE}}) \sim \mathcal{N}(0, \sigma^2/2)$. 
\end{enumerate}

Generating a set of $M$ statistically independent random matrices of identical dimension $L$ 
yields the standard ensemble denoted by $\text{GOE}(M, L)$. This sample size allows for the 
computation of bootstrapped confidence intervals for all evaluated spectral observables \cite{efron94, davison97}. 
While confidence intervals are rarely utilized in standard random matrix theory (RMT) studies, 
they are essential in the present context to quantify finite-sample fluctuations inherent to the 
numerical generation of mixed ensembles.

A mixed ensemble is constructed by varying the matrix dimensions within a sample of size $M$.
This mixed Gaussian Orthogonal Ensemble, designated as $\text{mGOE}(M, L, \mu)$, is parameterized 
by the asymmetry parameter $\mu$, which regulates the variance of the underlying size distribution. 
In the deterministic limit where $\mu \to 1.0$, the matrix size fluctuations vanish, recovering 
the conventional GOE exactly: $\text{GOE}(M, L) = \lim_{\mu \to 1.0} \text{mGOE}(M, L, \mu)$.

The numerical generation protocol for the $\text{mGOE}(M, L, \mu)$ proceeds as follows:
\begin{enumerate}
\item A sequence of matrix sizes $\{n_i\}_{i=1}^M$ is generated, where each dimension $n_i$ 
is drawn from a binomial distribution. Specifically, the deviation from a base dimension is regulated 
by $L$ independent Bernoulli trials with a success probability $\mu$, establishing a mapping 
where $n_i$ represents a realization of a scaled binomial process.
\item For each generated dimension $n_i$, a random matrix is sampled independently from the 
corresponding $\text{GOE}(n_i)$.
\item The degree of mixture $\mu$ acts as the controlling parameter of this stochastic process. As 
$\mu \to 1.0$, the dimension assignment becomes deterministic, forcing $n_i = L$ for all $i$, 
which matches the conventional homogeneous ensemble layout.
\item In the large-$M$ limit, the ensemble satisfies the macroscopic consistency 
condition $L \mu = \frac{1}{M} \sum_{i=1}^M n_i$, ensuring that the mean matrix dimension 
converges robustly to its analytically expected value.
\end{enumerate}

\section{Spectral Periodicity and the Composite System Mapping} \label{app:per}

The structural connection between the mGOE framework and the composite system arises 
because the mGOE systematically samples a distribution of matrix dimensions $n_i$ clustered around 
the sub-selection scale. To enable a meaningful statistical comparison across different matrix sizes, 
the respective spectra must be aligned by imposing periodic boundary conditions. For an individual 
matrix realization of size $n_i$, the spectrum yields $n_i$ unique eigenvalues. 

Periodicity enforces a cyclic repetition of these eigenvalues up to the reference cutoff dimension $N$, 
consistent with the structural requirements of the $\text{mGOE}(M, L, \mu)$ model. This wrapping protocol 
introduces exact spectral degeneracies that accurately mirror the subspace sub-selection procedure of 
the composite Hamiltonian at a statistical level. The parameter $\mu$ governs this continuous crossover, 
reproducing the non-scaling localization phenomenon observed in the underlying physical states. Beyond its 
direct utility in analyzing non-ergodic localization transitions, the mGOE framework offers a general model 
applicable to the study of spectral fluctuations in disordered complex networks and quantum graphs.

\section{Analytical Onset of Spectral Bimodality}
\label{app:bimodality}

To provide an analytical foundation for the emergence of the bimodal spectral profile characteristic 
of the Wigner cat phases, we examine the spectral moments of the uncoupled block-diagonal operator:

\begin{equation}
H = \begin{pmatrix} H^{(1)} & \mathbf{0} \\ \mathbf{0} & H^{(2)} \end{pmatrix},
\end{equation}
where $H^{(1)}$ and $H^{(2)}$ are independent, real-symmetric GOE matrices of dimension 
$N_1 \times N_1$ and $N_2 \times N_2$, respectively, defining a total Hilbert space dimension 
of $N = N_1 + N_2$. The normalized kurtosis ratio, defined as $\kappa = M_4 / (M_2)^2$ where 
$M_k = \frac{1}{N} \mathbb{E}[\text{Tr}(H^k)]$, serves as a reliable macroscopic signature 
characterizing the global morphology of the spectral density \cite{mehta04}.

Under a global variance scaling constraint ($\sigma^2 = 1/N$), the linear additivity 
of the independent block traces establishes that $M_2 = \frac{1}{N} ( \mathbb{E}[\text{Tr}(H^{(1)})^2] + 
\mathbb{E}[\text{Tr}(H^{(2)})^2] ) \longrightarrow 1$ in the asymptotic limit $N \to \infty$. 

Conversely, expanding the fourth spectral moment isolates the contribution of closed-loop random walks 
of length four ($i \to j \to k \to l \to i$). Because the off-diagonal coupling blocks are strictly zero, 
any path attempting to bridge the decoupled subspaces ($i \in N_1$ and $k \in N_2$) is identically 
eliminated. This explicit \textit{off-block path suppression} simplifies the combinatorial path counting 
to a strict function of the sub-block macro-dimensions \cite{mehta04}:

\begin{equation}
M_4 = \frac{3}{N} \left( \frac{N_1^3}{N} + \frac{N_2^3}{N} \right) = 3 \left( 1 - \frac{2N_1 N_2}{N^2} \right).
\end{equation}
Combining these moment equations yields the simplified expression for the spectral kurtosis:
\begin{equation}
\kappa = 3 \left( 1 - \frac{2N_1 N_2}{N^2} \right).
\end{equation}

This structural expression provides analytical verification that spectral bimodality emerges directly 
from spatial state-space partitioning. In the trivial limit where the system collapses into a single 
unbroken domain ($N_1 \to N, N_2 \to 0$), the expression recovers the standard Gaussian/unimodal benchmark 
$\kappa = 3.0$ \cite{mehta04}. However, for any non-trivial block partitioning ($N_1, N_2 > 0$), the 
suppression of cross-subspace random walks drives $\kappa$ strictly below this threshold, reaching a minimum 
value of $\kappa = 1.5$ at a perfectly symmetric split ($N_1 = N_2 = N/2$). Because a sub-Gaussian kurtosis 
($\kappa < 3.0$) indicates that spectral mass is displaced from the origin toward the outer boundaries, this 
central valley collapse mathematically guarantees the onset of the bimodal Wigner cat phase directly from the 
composition rule, independent of post-processing spectral truncation.

\end{document}